\documentclass[%
 reprint,
 amsmath,amssymb,
 aps,
pra,
]{revtex4-1}

\usepackage{graphicx}
\usepackage{dcolumn}
\usepackage{bm}
\usepackage{physics}
\usepackage{subcaption}
\usepackage{braket}
\usepackage{amssymb}
\usepackage{float}
\usepackage[flushleft]{threeparttable}

\begin{document}

\preprint{APS/123-QED}

\title{Squeezed state evolution and entanglement in lossy coupled resonator optical waveguides}
\author{Hossein Seifoory}
\email{hossein.seifoory@queensu.ca}
\author{Marc. M. Dignam}
 \affiliation{Department of Physics, Engineering Physics and Astronomy, Queen's University, Kingston, Ontario K7L 3N6, Canada 
}%

\date{\today}

\begin{abstract}
We investigate theoretically the temporal evolution of a squeezed state in lossy coupled-cavity systems. We present a general formalism based upon the tight binding approximation and apply this to a two-cavity system as well as to a  coupled resonator optical waveguide in a photonic crystal. We derive analytical expressions for the number of photons and the quadrature noise in each cavity as a function of time when the initial excited state is a squeezed state in one of the cavities. We also analytically evaluate the time dependant cross correlation between the photons in different cavities to evaluate the degree of quantum entanglement. We demonstrate the effects of loss on the properties of the coupled-cavity systems and derive approximate analytic expressions for the maximum photon number, maximum squeezing and maximum entanglement for cavities far from the initially excited cavity in a lossless coupled resonator optical waveguide.        
\end{abstract}

\pacs{Valid PACS appear here}
\maketitle


\section{\label{sec:level1}Introduction}

Nonclassical states of light possess properties that can only be described by quantum theory. One potential attribute of these states is quantum entanglement, which has potential applications in quantum teleportation, quantum computation, and quantum information~\cite{cerf2007quantum,bouwmeester2013physics}. Both discrete-variables (DVs) and continuous-variables (CVs) can be used to create quantum entanglement between two distant quantum systems. However, implementation of DV entanglement currently suffers from difficulties in single photon generation and detectionand from loss in integrated on-chip systems. In contrast, CV entanglement as an alternative to its DV counterpart can be efficiently created and used for implementation of CV quantum protocols~\cite{Huang2016,PhysRevA.83.042312,PhysRevX.5.041010,PhysRevLett.93.250503,PhysRevA.80.050303} and has the advantage that the entanglement is general more robust to loss than systems composed of photon pairs. \par  
Generally, non-uniformly distributed quadrature fluctuations of squeezed light can provide CV entanglement~\cite{PhysRevLett.84.3482,RevModPhys.77.513}. The inseparability criterion, which is based on the total variance of a pair of canonical conjugates variables, can be used to study the degree of quantum correlation in CV systems~\cite{PhysRevLett.84.2722,PhysRevLett.84.2726}. Although the CV entangled light has been achieved using bulk setups~\cite{PhysRevLett.93.250503,PhysRevLett.84.3482,Marino2009,PhysRevA.76.053827,PhysRevLett.98.240401}, the migration from bulk optics to integrated photonics seems inevitable since, as the size and complexity of these systems increase,
the limitations of working with the bulk optics, such as stability, precision, and physical size,
become significant. Due to recent developments in integrated photonics technology, it is possible to resolve scalability and stability concerns regarding to bulk optics by generating CV entanglement on a chip~\cite{Masada2015}. However, the effects of environmental loss, which destroys the nonclassical properties of light and consequently affects the entanglement~\cite{PhysRevA.85.052330}, is inevitable and needs to be understood and managed. \par
The coupled resonator optical waveguide (CROW), which was first studied by Yariv \textit{et. al} ~\cite{Yariv:99}, has been shown to have potential in generating CV entangled states between two spatially separated sites~\cite{PhysRevA.83.062310}. It can also be integrated with other photonic components, forming an integrated photonic circuit for use in photonic quantum information processing. 
In general, the CROW structure can be described as a waveguide consisting of weakly coupled optical cavities along one-dimension. The tight-binding method~\cite{ashcroft2011solid, doi:10.1063/1.2737430}, which uses localized single-cavity modes as a basis, can be applied as a mathematical framework to model the evolution of light in such a coupled structure. One nice feature of CROWs is that, by adjusting the nature of the cavities and the separation between the cavities, one can adjust the dispersion and even the loss to some degree to optimize the system for a particular application~\cite{Yariv:99}. This characteristic is the main advantage of using CROWs compared to conventional optical waveguides, in which the guiding properties are mostly determined by total internal reflection and material dispersion.\par
\begin{figure}
\begin{subfigure}{0.5\textwidth}
\includegraphics[width=\linewidth]{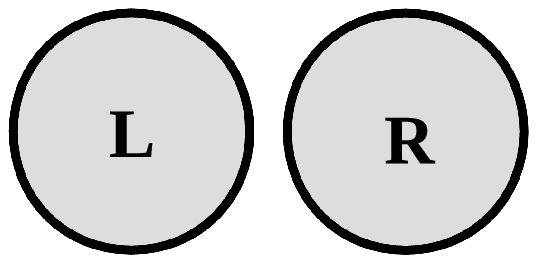}
\caption{} \label{fig:1a}
\end{subfigure}
\begin{subfigure}{0.49\textwidth}
\includegraphics[width=\linewidth]{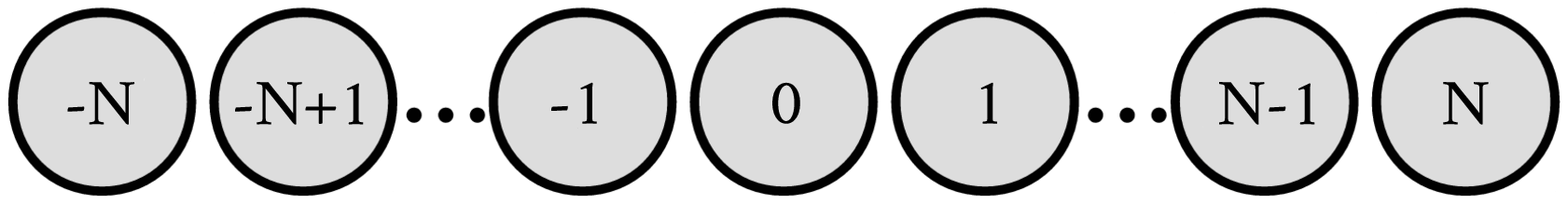}
\caption{} \label{fig:1b}
\end{subfigure}
\begin{subfigure}{0.49\textwidth}
\includegraphics[width=\linewidth]{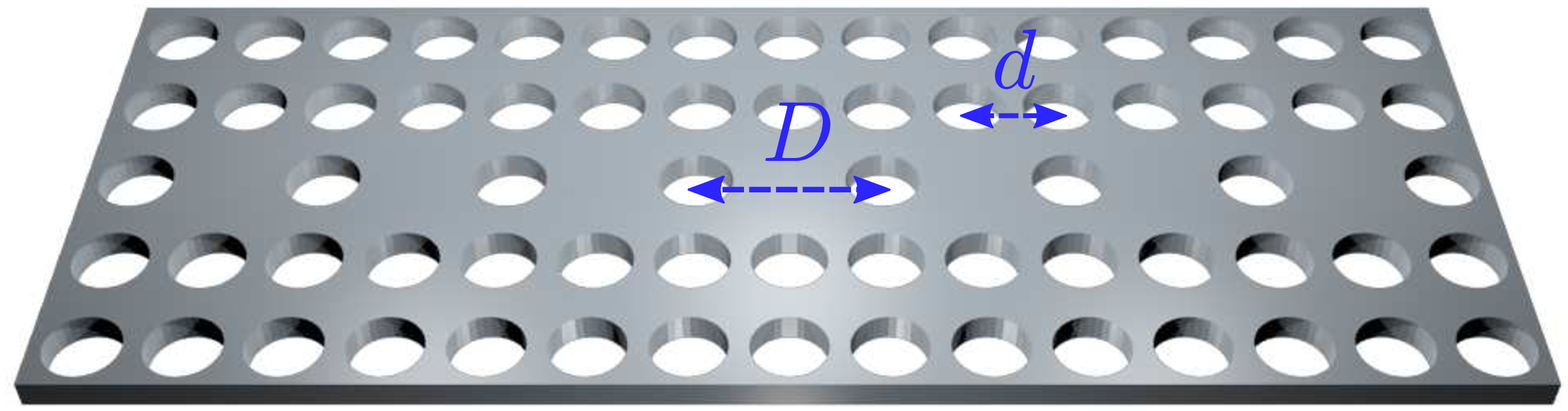}
\caption{} \label{fig:1c}
\end{subfigure}
\caption{Schematic picture of (a) two coupled cavities and (b) a CROW structure and (c) the particular CROW structure with period $D$ formed from defects in a slab photonic crystal with a square lattice of period $d$.}
\label{fig:schematic}
\end{figure}
In this paper, we show that the time evolution of the quadrature variances and the number of photons in each cavity can be explained and parametrized using a tight-binding approach. This method allows us to calculate the CV correlation variance between the photons in different cavities. The full analytic study of the squeezed state evolution in a lossy coupled-cavity system can provide us with insight into the influence of coupling and loss on the photon statistics and the nonclassical properties of the photons inside each cavity. We present the analytic expressions for a general initial state, but only explicitly present detailed results for an initial state which is a squeezed vacuum state in one of the cavities. \par  
The paper is organized as follows. In Sec.~\ref{sec:general}, we present the tight-binding formalism that is used to obtain the quasimodes~\cite{PhysRevA.77.053805} of the coupled-cavity system. We then derive the time dependent equations for the number of photons, the quadrature noise, and the CV correlation variance in a lossy coupled-cavity system with a general initial state. In Sec.~\ref{sec:two_cavity}, as a test system, we study the state evolution in a lossy two-coupled-cavity system (see Fig. \ref{fig:1a}), where the initial state is a squeezed state in one of the cavities. In Sec.~\ref{sec:CROW}, we apply our approach to examine the same quantities for the state in the more technologically-interesting CROW structure (see Figs. \ref{fig:1b} and \ref{fig:1c}), where the initial state is also a squeezed state in one of the cavities. Finally, in Sec.~\ref{sec:conclusion}, we present our conclusions.    
\section{General formalism}
\label{sec:general}
In this section, we first present the general form of tight binding theory and derive the general expressions for some important quantities such as the time evolution of photon number, variances of quadrature operators, and correlation variance in lossy coupled-cavity systems.   \par
Although in this paper we mainly focus on the Squeezed vacuum state~(SVS) as the initial state of the $c^{th}$ cavity, in this section we consider a more general case and present analytic results for a general initial state.\par
Using the tight-binding formalism~\cite{doi:10.1063/1.4897523,doi:10.1063/1.2737430}, we can determine the fields and complex frequencies for the leaky modes of a coupled-cavity system. This formalism allows us to determine the mode fields and frequencies of a lossy coupled-cavity structure using only one finite-difference time domain (FDTD) calculation.\par   
The modes of a system can be obtained by solving the corresponding homogeneous Helmholtz equation for the electric field. However, here, due to the leakage in the system, we employ \textit{quasimodes} (QMs) which are electromagnetic resonances of an open~(leaky) dielectric structure and are characterized by complex frequencies, $\tilde{\omega}_m$. We denote the complex mode field of these QMs by $\tilde{\mathbf{N}}_m (\mathbf{r})$.  \par 
Following Fussel and Dignam~\cite{doi:10.1063/1.2737430}, we begin by expanding the coupled-cavity QMs, $\tilde{\mathbf{N}}_m (\mathbf{r})$, in terms of single-cavity QMs, $\tilde{\mathbf{M}}_q (\mathbf{r})$, as
\begin{equation}
\tilde{\mathbf{N}}_m(\mathbf{r})=\sum_{q}^{}v_{mq}\tilde{\mathbf{M}}_q (\mathbf{r}),
\label{eq:expansion}
\end{equation}
where $q$ labels the mode associated with a given cavity, $m$ labels a given coupled-cavity mode and $v_{mq}$ are the expansion coefficients.
The quasimodes are the solutions to the homogeneous Helmholtz equation for the electric field in the coupled-cavity and single-cavity structures:
\begin{equation}
\curl{\curl{\tilde{\mathbf{N}}_m (\mathbf{r})}}-\frac{\tilde{\omega}_{m}^2}{c^2}\epsilon(r)\tilde{\mathbf{N}}_m (\mathbf{r})=0,
\label{eq:helmholtz_epsilon}
\end{equation} 
\begin{equation}
\curl{\curl{\tilde{\mathbf{M}}_q (\mathbf{r})}}-\frac{\tilde{\Omega}_{q}^2}{c^2}\epsilon_{q}(\mathbf{r})\tilde{\mathbf{M}}_q (\mathbf{r})=0,
\label{eq:helmholtz_epsilon_q}
\end{equation} 
where $\tilde{\omega}_m \equiv \omega_m - i\gamma_m$ is the complex frequency for the $m^{th}$ QM of the coupled-cavity structure and $\tilde{\Omega}_q \equiv \Omega_q - i\Gamma_q$ is the complex frequency of the $q^{th}$ cavity. Also, $\epsilon(\mathbf{r})$ and $\epsilon_q(\mathbf{r})$ are the dielectric material profiles of the full coupled-cavity structure and the structure that only contains the $q^{th}$ cavity, respectively. The single-cavity modes and frequencies are calculated using FDTD.\par 
Substituting Eq.~(\ref{eq:expansion}) into Eq.~(\ref{eq:helmholtz_epsilon}) and then using Eq.~(\ref{eq:helmholtz_epsilon_q}) leads to the generalized eigenvalue equation,
\begin{equation}
\tilde{\mathbf{A}}\tilde{\mathbf{\Omega}} \tilde{\mathbf{v}}=\tilde{\mathbf{\Lambda}}(\tilde{\mathbf{A}}+\tilde{\mathbf{B}})\tilde{\mathbf{v}},
\label{eq:eigenvalue}
\end{equation}
where $\tilde{\mathbf{\Lambda}}\equiv Diag{\{\tilde{\omega}_m^2\}}$, $\tilde{\mathbf{\Omega}}\equiv Diag{\{\tilde{\Omega}_q^2\}}$ and $\tilde{\mathbf{v}}\equiv \{\tilde{v}_{mq}\}$ and $\tilde{\mathbf{A}}$ and $\tilde{\mathbf{B}}$ are respectively the overlap and coupling coefficients between $p^{th}$ and $q^{th}$ cavities with elements defined as
\begin{equation}
\tilde{A}_{qp}=\int d^{3}\textbf{r} ~\epsilon_{q}(\textbf{r})~\tilde{\textbf{M}}_q^*(\textbf{r})\cdot\tilde{\textbf{M}}_p(\textbf{r}),
\label{A}
\end{equation}   
\begin{equation}
\tilde{B}_{qp}=\int d^{3}\textbf{r}~ \delta \epsilon_{q}(\textbf{r})~\tilde{\textbf{M}}_q^*(\textbf{r})\cdot\tilde{\textbf{M}}_p(\textbf{r}),
\label{B}
\end{equation}
where $\delta \epsilon_{q}(\textbf{r})\equiv \epsilon(\textbf{r})-\epsilon_{q}(\textbf{r})$.\par 
Using the expansion coefficients, $v_{mq}$, the annihilation operator, $b_m$, for the $m^{th}$ mode of the coupled-cavity system can be written in terms of the $q^{th}$ individual single-mode cavity operators, $a_q$, as
\begin{equation}
b_m=\sum_{q}^{}\tilde{v}^*_{mq}a_{q}.
\label{eq:expansion_b_a}
\end{equation}\par 
 Although we will be interested in the nature of the states in individual cavities, the evolution is most simply calculated using the full coupled-cavity annihilation operators. This evolution is found by solving the adjoint master equation for this open, lossy system~\cite{open_quantum_system}. We have previously shown that for any product of normally ordered operators, the time dependence of the individual annihilation operators is given by
 \begin{equation} 
 b_m(t)=b_m e^{-i\tilde{\omega}_m t},
 \label{evol_b_m}
 \end{equation}
where $b_m=b_m(0)$ is the corresponding operator in the Schrodinger representation~\cite{PhysRevA.85.013809} and the time evolution of the creation operator is simply the Hermitian conjugate of Eq.~(\ref{evol_b_m}).\par
Using Eqs.~(\ref{eq:expansion_b_a}) and (\ref{evol_b_m}), one can generally write the time-dependant localized field operator in a coupled-cavity system in terms of expansion coefficients and the operators at $t=0$ as
\begin{equation}
a_p(t)=\sum_{mq}^{}\tilde{v}^*_{mq}a_{q}e^{-i\tilde{\omega}_m t} \tilde{v}_{mp}.
\label{eq:general_ap_t}
\end{equation}
Using Eq.~(\ref{eq:general_ap_t}) and its complex conjugate, the time dependent average photon number in the $p^{th}$ cavity can be written as
\begin{equation}
\braket{a^\dagger_p(t)a_p(t)}=
\sum_{mqm'q'}\braket{a^\dagger_q a_{q'}}\tilde{v}^*_{mp}\tilde{v}_{mq}\tilde{v}^*_{m'q'}\tilde{v}_{m'p}e^{-i(\tilde{\omega}_{m'}-\tilde{\omega}_m^*)t}.
\label{eq:general_photon_number}
\end{equation} 
Following the same procedure for the quadrature operators, $X_p=a_p+a_p^{\dagger}$ and $Y_p=i(a_p-a_p^{\dagger})$, the time-dependent variances of $X$ quadrature operator in general form can be shown to be
\begin{equation}
\label{eq:general_DeltaX_CROW}
\begin{split}
\braket{(\Delta X_p)^2}&=1+\\\sum_{mqm'q'}\Big(
&2\braket{a^\dagger_q a_{q'}}\tilde{v}^*_{mp}\tilde{v}_{mq}\tilde{v}^*_{m'q'}\tilde{v}_{m'p}e^{-i(\tilde{\omega}_{m'}-\tilde{\omega}_m^*)t}\\
&+\braket{a_q a_{q'}}\tilde{v}^*_{mq}\tilde{v}_{mp}\tilde{v}^*_{m'q'}\tilde{v}_{m'p}e^{-i(\tilde{\omega}_{m'}+\tilde{\omega}_m)t}\\
&+\braket{a^\dagger _q  a^\dagger _{q'}}\tilde{v}^*_{mp}\tilde{v}_{mq}\tilde{v}^*_{m'p}\tilde{v}_{m'q'}e^{i(\tilde{\omega}^*_{m'}+\tilde{\omega}^*_m)t}\Big).
\end{split}
\end{equation} 
\par
To obtain the general form of $\braket{(\Delta Y_p)^2}$, one just needs to change the sign of the last two terms in Eq.~(\ref{eq:general_DeltaX_CROW}), containing $\braket{a_q a_{q'}}$ and $\braket{a^\dagger _q  a^\dagger _{q'}}$.\par
We now study the degree of entanglement between the photons in cavities $p$ and $p'$ in a coupled-cavity structure using correlation variance, which is defined as~\cite{PhysRevLett.84.2722,PhysRevLett.84.2726}
\begin{equation}
\Delta_{p,p'}^2=\braket{[\Delta(X_p-X_{p'})]^2}{}+\braket{[\Delta(Y_p+Y_{p'})]^2}{}.
\label{eq:correlation_nm_def}
\end{equation} 
Employing Eq.~(\ref{eq:general_ap_t}) and its complex conjugate in Eq.~(\ref{eq:correlation_nm_def}), the general form of the time-dependent correlation variance  can be written as
\begin{widetext}
\begin{equation}
\begin{split}
\Delta_{p,p'}^2=&4+4\sum_{mqm'q'}\Big(\braket{a^\dagger_q a_{q'}}\big(\tilde{v}^*_{mp}\tilde{v}_{mq}\tilde{v}^*_{m'q'}\tilde{v}_{m'p}+\tilde{v}^*_{mp'}\tilde{v}_{mq}\tilde{v}^*_{m'q'}\tilde{v}_{m'p'}\big)e^{-i(\tilde{\omega}_{m'}-\tilde{\omega}_m^*)t}\Big)\\
&-4\sum_{mqm'q'}\braket{a_q a_{q'}}\tilde{v}^*_{mq}\tilde{v}_{mp}\tilde{v}^*_{m'q'}\tilde{v}_{m'p'}e^{-i(\tilde{\omega}_{m'}+\tilde{\omega}_m)t}-4\sum_{mqm'q'}\braket{a^\dagger _q  a^\dagger _{q'}}\tilde{v}^*_{mp}\tilde{v}_{mq}\tilde{v}^*_{m'p'}\tilde{v}_{m'q'}e^{i(\tilde{\omega}^*_{m'}+\tilde{\omega}^*_m)t}.
\end{split}
\label{eq:general_entanglement}
\end{equation}
\end{widetext}
Equations (\ref{eq:general_photon_number}), (\ref{eq:general_DeltaX_CROW}) and (\ref{eq:general_entanglement}) give the results for general initial conditions.  In the rest of this paper, we  assume that at time $t=0$, the $c^{th}$ cavity is in an excited state of light and the rest of the cavities are in the vacuum state; thus, in what follows, the only nonvanishing terms in Eqs.~(\ref{eq:general_photon_number}), (\ref{eq:general_DeltaX_CROW}), and (\ref{eq:general_entanglement}) are those in which $q=q'=c$. Before proceeding, we first briefly review some of the quantum properties of different initial states in a single cavity. In Table~\ref{table1} we summarize some properties of three different initial states: the SVS, squeezed thermal state~(STS), and coherent state. In Table~\ref{table1}, $\tilde{\eta}$ is the coherent state parameter, and $n_{th}$ is the thermal photon number for the STS. For the SVS and STS, the squeezing parameter is generally complex, and we write it in the form $ \tilde{\xi} = u e^{i\phi} $, where $u$ and $\phi$ are the squeezing amplitude and phase, respectively.\par 
The formalism introduced here is used in the following sections to study the temporal evolution of a SVS, first in a simple two coupled-cavity system and then in a CROW structure.\par 
\setlength{\tabcolsep}{0.6em}
\begin{table*}
 \begin{threeparttable}
\centering
\caption{The operator expectation values for the SVS, STS, and coherent state for the $c^{th}$ cavity at $t=0$. Here, $u$ and $\phi$ are the squeezing amplitude and phase, respectively, for the SVS and STS, $\tilde{\eta}$ is the coherent state parameter, and $n_{th}$ is the thermal photon number for the STS.} 
\label{table1}
 \begin{tabular}{c  c  c  c} 
 \hline \hline\\
  & SVS & STS & Coherent state \\ [1ex] 
 \hline\\
 $\braket{a^\dagger _c  a_{c}}$ & $\sinh[2](u)$ & $n_{th}\cosh(2u)+\sinh[2](u)$ & $|\tilde{\eta}|^2$\\ [1ex]
 $\braket{a_c  a_{c}}$& $-e^{i\phi}\cosh(u)\sinh(u)$ & $-(n_{th}+\frac{1}{2})e^{i\phi}\sinh(2u)$  & ${\tilde{\eta}}^2$  \\[1ex]
 $\braket{a^\dagger _c  a^\dagger_{c}}$ & $-e^{-i\phi}\cosh(u)\sinh(u)$ & $-(n_{th}+\frac{1}{2})e^{-i\phi}\sinh(2u)$  & ${\tilde{\eta}^{*2}}$\\
 [1ex]
 \hline \hline
 \end{tabular}
 \end{threeparttable}
\end{table*}
\section{Two lossy coupled-cavities}
\label{sec:two_cavity}
We first consider a system consisting of only two identical lossy coupled cavities, shown in Fig.~(\ref{fig:1a}). This system supports two QMs, $\tilde{\omega}_+$ and $\tilde{\omega}_-$ representing the symmetric and antisymmetric QMs, respectively. It has been shown that in the NNTB approximation, the supporting QMs of a two coupled-cavity system can be written in terms of coupling parameters~\cite{1017595,Kamalakis} as
\begin{equation}
\tilde{\omega}_{\pm}\simeq\tilde{\Omega}_0(1\pm\tilde{\beta}_1/2),
\end{equation}
where $\tilde{\beta}_1\equiv\tilde{B}_{LR}$. The general quantum states of such a system can be expanded in terms of the two individual cavity states as
\begin{equation}
\tilde{\textbf{M}}_\pm(\textbf{r})=\frac{1}{\sqrt{2}}(\tilde{\textbf{N}}_L(\textbf{r}) \pm \tilde{\textbf{N}}_R(\textbf{r})),
\end{equation} 
where $\tilde{\textbf{N}}_L(\textbf{r}$) and $\tilde{\textbf{N}}_R(\textbf{r})$ are the modes of the left and right cavities, respectively. Using Eq.~(\ref{eq:expansion_b_a}) with $v_{+L}=v_{+R}=1/\sqrt{2}$ and $v_{-L}=-v_{-R}=1/\sqrt{2}$, one can also write the coupled-cavity operators as symmetric and antisymmetric superpositions of the localized site operators as
\begin{equation}
b_{+}=\frac{1}{\sqrt{2}}(a_L+a_R)
\label{b_plus}
\end{equation}
and
\begin{equation}
b_{-}=\frac{1}{\sqrt{2}}(a_L-a_R),
\label{b_minus}
\end{equation}
where $a_L$ and $a_R$ are the annihilation operators acting on the left and right cavities, respectively.\par
To examine the propagation and time evolution of nonclassical light in lossy coupled-cavity structures, we begin by studying the evolution of squeezed light in our lossy two-cavity system. Using Eqs.~(\ref{eq:general_photon_number}), (\ref{eq:general_DeltaX_CROW}), and (\ref{eq:general_entanglement}), we study the time evolution of the quadrature variances and photon statistic of states in both cavities to investigate whether the light transferred to the second cavity  maintains its nonclassical properties and to determine the correlations between light in the different cavities. \par 
In all the following equations, the specific state of the $c^{th}$ cavity at t=0 is general. Table~\ref{table1} can be used to obtain results for specific initial states. However, all plotted results in this section will be for the initial state where the $c^{th}$ cavity is the left cavity, which is in a SVS with $u=1.2$ and $\phi=0$. We set $\tilde{\omega}_{\pm}=\omega_{\pm}-i\gamma_{\pm}$ and for simplicity, we assume that $\gamma_{+}=\gamma_{-}=\gamma$ and define $\omega_{+}=\omega$ and $\omega_{+}-\omega_{-}=\Delta$. We choose $\Delta=\omega/20$ and $\gamma=0.02\Delta$. The parameters are chosen to demonstrate some of the key features which are the beating back and forth between the two cavities, the free oscillations, and the loss in the system.\par 
Using Eqs.~(\ref{eq:general_photon_number}), the time-dependant average number of photons in the cavities can then be written as
\begin{equation}
\label{eq:two_num_R_0}
\begin{split}
\braket{a^{\dagger}_{R}(t)a_{R}(t)}{}=&
\frac{1}{4}\braket{a^{\dagger}_{L}a_L}{}(e^{i(\tilde{\omega}^*_+ -\tilde{\omega}_+)t}-e^{i(\tilde{\omega}^*_+ -\tilde{\omega}_-)t}\\
&-e^{i(\tilde{\omega}^*_- -\tilde{\omega}_+)t}+e^{i(\tilde{\omega}^*_- -\tilde{\omega}_-)t}),
\end{split}
\end{equation}
\begin{equation}
\label{eq:two_num_L_0}
\begin{split}
\braket{a^{\dagger}_{L}(t)a_{L}(t)}{}=&
\frac{1}{4}\braket{a^{\dagger}_{L}a_L}{}(e^{i(\tilde{\omega}^*_+ -\tilde{\omega}_+)t}+e^{i(\tilde{\omega}^*_+ -\tilde{\omega}_-)t}\\
&+e^{i(\tilde{\omega}^*_- -\tilde{\omega}_+)t}+e^{i(\tilde{\omega}^*_- -\tilde{\omega}_-)t}),
\end{split}
\end{equation}
which can be simplified as
\begin{equation}
\label{eq:two_num_R}
\braket{a^{\dagger}_{R}(t) a_{R}(t)}{}=\frac{1}{2}\braket{a^{\dagger}_{L} a_{L}}e^{-2\gamma t}[1-\cos(\Delta t)]
\end{equation}
and
\begin{equation}
\label{eq:two_num_L}
\braket{a^{\dagger}_{L}(t) a_{L}(t)}=\frac{1}{2}\braket{a^\dagger_L a_L}e^{-2\gamma t}[1+\cos(\Delta t)].
\end{equation}
As expected for lossy systems, at large times, the number of photons in both cavities decays to zero at long times, due to the exponential decay coefficient in Eqs.~(\ref{eq:two_num_R}) and (\ref{eq:two_num_L}).   \par
\begin{figure}
         
                \includegraphics[scale=0.44]{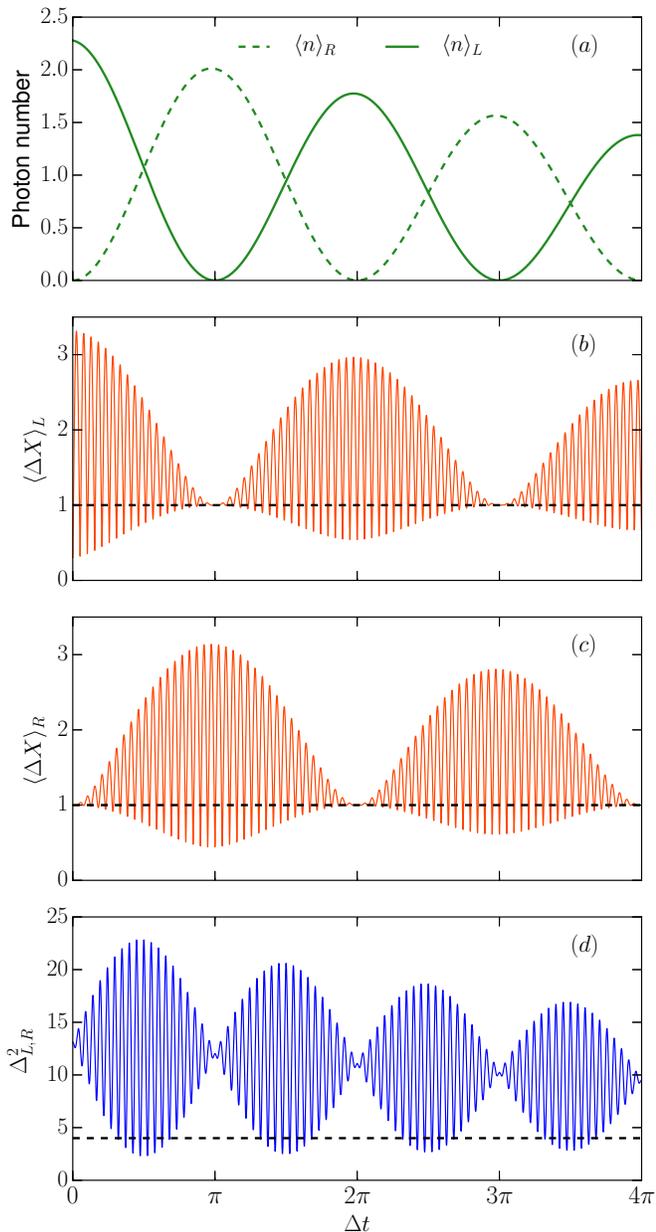}

                \caption{(Color online) Calculated results for the two coupled-cavity system. (a) Time evolution of the mean photon number in the left (solid) and right (dashed) cavities. The time evolution of the quadrature noise in $X$ in~(b) the left and~(c) right cavities. The dashed line at $\braket{\Delta X}_{(L,R)}=1$ shows the classical limit for the quadrature noises. In (d), we plot the CV correlation variance as a function of time. The dashed line in (d) shows the inseparability limit below which the light is considered to be entangled.} 
                \label{fig:n_DeltaX_LR_ent}
                
                \end{figure}

Following the same procedure as before and using Eq.~(\ref{eq:general_DeltaX_CROW}), the variances of quadrature operators can be shown to be
\begin{equation}
\label{eq:deltaX_LR}
\begin{split}
\braket{\Delta X^2}_{(L,R)}=&1+e^{-2\gamma t}(1\pm\cos(\Delta t))\Big(\braket{a^{\dagger}_L a_L}\\
&\pm\frac{1}{2}\braket{a_L a_L}e^{-i(2\omega-\Delta)t}\\
&\pm\frac{1}{2}\braket{a^{\dagger}_L a^{\dagger}_L}e^{i(2\omega-\Delta)t}\Big)
\end{split}
\end{equation}
and
\begin{equation}
\label{eq:deltaY_LR}
\begin{split}
\braket{\Delta Y^2}_{(L,R)}=&1+e^{-2\gamma t}(1\pm\cos(\Delta t))\Big(\braket{a^{\dagger}_L a_L}\\
&\mp\frac{1}{2}\braket{a_L a_L}e^{-i(2\omega-\Delta)t}\\
&\mp\frac{1}{2}\braket{a^{\dagger}_L a^{\dagger}_L}e^{i(2\omega-\Delta)t}\Big),
\end{split}
\end{equation} 
where the upper (lower) signs belong to the left (right) cavity. In Fig.~\ref{fig:n_DeltaX_LR_ent} (a) we plot the mean photon number in each cavity as it evolves in time for a SVS. 
As can be seen, the photons, which are all initially in the left cavity, periodically move in time between the two cavities. It is also evident that the mean photon number gradually decreases due to the scattering loss. The time dependant quadrature noise in $X$ is shown in Figs.~\ref{fig:n_DeltaX_LR_ent} (b) and (c)  for the left and right cavity, respectively. The dashed lines indicate the classical limit, below which the quadrature noise is squeezed. As expected, the quadrature noise in the left cavity is initially less than this limit~($\braket{\Delta X(t=0)}_{L}=0.3$) since it is a SVS. On the other hand, the right cavity at $t=0$ is vacuum and consequently has the minimum classical quadrature noise, $\Delta X_{R}=1$. As the coupled system evolves in time, we can see that the $\Delta X_{R}$ falls below the classical limit and reaches $0.4$ at $\Delta t=\pi$, which confirms that the squeezed state has been transferred to the right cavity due to the coupling between the two cavities. In the absence of loss, this squeezing would be identical to the squeezing in the left cavity at $t=0$.\par
Before moving to the CROW structure, we evaluate the entanglement between the light in these two cavities. It has been shown that $ \Delta_{L,R}^2 <4$ can be considered as the inseparability criterion for entanglement~\cite{PhysRevLett.84.2722,PhysRevLett.84.2726,Masada2015,Zhang2015}. Using Eq.~(\ref{eq:general_entanglement}) and considering the same initial condition as before, the time-dependant correlation variance is found to be 
\begin{equation}
\label{eq:corr_var_two}
\begin{split}
\Delta^2_{L,R}=&4+e^{-2\gamma t}\Big(4\braket{a^\dagger_L a_L}\\
&-\braket{a_L a_L}e^{i\Delta t}(e^{-i(2\omega+\Delta)t}-e^{-i(2\omega-\Delta)t})\\
&-\braket{a^\dagger_L a^\dagger_L}e^{-i\Delta t}(e^{+i(2\omega+\Delta)t}-e^{+i(2\omega-\Delta)t})  \Big).
\end{split}
\end{equation}
Using Eq.~(\ref{eq:corr_var_two}) for a lossless system and considering the initial excited state in the left cavity to be a SVS with large squeezing amplitude ($u\gg1$), it can be shown that the minimum achievable value of $\Delta^2_{L,R}$ is $2$, which is well below the inseparability limit. In Fig.~\ref{fig:n_DeltaX_LR_ent} (d) we plot the CV correlation variance as a function of time. As can be seen, the correlation variance exceeds the inseparability criterion reaching local minima close to the times, $\Delta t=(2l+1)\pi/2$, where $l$ is an integer. As expected, the loss affects the degree of inseparability as the system evolves in time. For instance, due to loss, although the correlation variance at $\Delta t=7\pi/2$ is still below the inseparability limit, it experiences about $22\%$ increase compared to the time $\Delta t=\pi/2$ where $\Delta^2_{L,R}\approx 2.3$.   

\section{Coupled resonator optical waveguides}
\label{sec:CROW}
Although the simple two-coupled-cavity system is a useful testbench for understanding the evolution of different states of light in lossy coupled systems, to be more practical, we examine a CROW structure, in which the light can propagate over a longer distance. Such a system, consisting of $2N+1$ weakly coupled lossy optical cavities along one-dimension with a periodicity $D$, is schematically shown in Fig.~\ref{fig:schematic}~(b). As discussed in Sec.~\ref{sec:general}, this system can be studied using the tight-binding method which assumes weak coupling between different cavities and uses localized single mode cavity as a basis.\par
Assuming that all the cavities are identical and support the same mode, $\tilde{\Omega}_p=\tilde{\Omega}_0$, and using the fact that $\tilde{A}_{pq}=\tilde{A}_{p~-q}$ and $\tilde{B}_{pq}=\tilde{B}_{p~-q}$ in Eq.~(\ref{eq:eigenvalue}), then applying periodic boundary condition and Bloch's theorem, the tight-binding dispersion can be written as
\begin{equation}
\tilde{\omega}(k)=\tilde{\Omega}_0\sqrt{\frac{1+2\sum\limits_{p=1}^{N}\cos(kpD)\tilde{\alpha}_{p}}{1+2\sum\limits_{p=1}^{N}\cos(kpD)(\tilde{\alpha}_{p}+\tilde{\beta}_{p})}},
\label{eq:dispersion}
\end{equation}
where $\tilde{\alpha}_p\equiv \tilde{A}_{0p}$ and $\tilde{\beta}_p \equiv \tilde{B}_{0p}$. Using the NNTB approximation, where only $\tilde{\beta}_1\neq 0$ in Eq.~(\ref{eq:dispersion}), we obtain
\begin{equation}
\tilde{\omega}(k)\approx\tilde{\Omega}_0[1-\tilde{\beta}_1 \cos(kD)],
\label{eq:dispersion_NNTB}
\end{equation}    
where we used the Taylor expansion of the square root function. It can be seen from Eq.~(\ref{eq:dispersion_NNTB}), the modes of the CROW experience different loss rates, which can differ by an order of magnitude~\cite{MohsenThesis,Fussell:07}. Again from NNTB we obtain 
\begin{equation}
\label{eq:v_kp}
v_{kp}=\frac{e^{ikpD}}{\sqrt{N}}.
\end{equation} \par
We take the $c^{th}$ cavity to be initially in a squeezed vacuum state, while all other cavities are in the vacuum state. Such a state could be achieved, for example, by strongly pumping the $c^{th}$ cavity in the presence of spontaneous parametric down-conversion~\cite{PhysRevLett.57.2520,introductory}, as long as the pump duration is much shorter than the time required for transfer to the neighbouring cavities.\par 
Employing Eqs.~(\ref{eq:general_photon_number}) and (\ref{eq:v_kp}), the time dependant average photon number in the $p^{th}$ cavity can be written as
\begin{equation}
\begin{split}
\braket{a^\dagger _p (t) a_p(t)}=&\frac{1}{N^2}\sum_{kqk'q'}{}\braket{a^\dagger _q  a_{q'}} (e^{-ik(p-q)D}e^{ik'(p-q')D}\\
&\cross e^{-i\tilde{\Omega}_0\left(1-\tilde{\beta}^*_1\cos\left(kD\right)\right)t}e^{-i\tilde{\Omega}_0\left(1-\tilde{\beta}_1\cos\left(k'D\right)\right)t}).
\label{eq:photon_number_CROW_0}
\end{split}
\end{equation} 
For our initial conditions, the only nonvanishing expectation value in Eq.~(\ref{eq:photon_number_CROW_0}) is $\braket{a^\dagger _c  a_{c}}=\sinh[2](u)$. Converting the sums to integrals and using the following equations
\begin{subequations}
\begin{align}
\int_{0}^{\pi}\cos(\tilde{z}\cos(x))\cos(nx) dx&=\pi\cos(\frac{n\pi}{2})J_n(\tilde{z}),\\
\int_{0}^{\pi}\sin(\tilde{z}\cos(x))\cos(nx) dx&=\pi\sin(\frac{n\pi}{2})J_n(\tilde{z}),
\label{eq:math_integral2}
\end{align}
\end{subequations}
where $n$ is an integer and $J_n(\tilde{z})$ is the Bessel function of the first kind of order $n$~\cite{Olver:2010:NHM:1830479}, one founds that the time dependant average photon number in the $p^{th}$ cavity is
\begin{equation}
\braket{a^\dagger _p (t) a_p(t)}=\braket{a^\dagger _c  a_{c}} e^{-2\gamma t} |J_{\delta p}(\tilde{\zeta}_1 t)|^2,
\label{eq:photon_number_CROW}
\end{equation} 
where $\tilde{\zeta}_1\equiv \tilde{\Omega}_0 \tilde{\beta}_1$, $\gamma=-\Im(\tilde{\Omega}_0)$, and $\delta p\equiv p-c$. To scale the time, we define $\tau=1/\Re{\tilde{\zeta}_1}$, which is the minimum time for a pulse to travel one period. In all of the plots in this section, it is assumed that the $c^{th}$ cavity is the one in the middle of the CROW structure~$(c=0)$ and it contains a SVS with $u=0.88$ and $\phi=0$, while the rest are initially in the vacuum state.\par
The physical parameters of the CROW considered in this paper are from Ref.~\cite{doi:10.1063/1.2737430}. The CROW consists of a dielectric slab of refractive index $n=3.4$ having a square array of cylindrical air void of radius $a=0.4d$, height $h=0.8d$, and lattice vectors $\textbf{a}_1=d\hat{\textbf{x}}$ and $\textbf{a}_2=d\hat{\textbf{y}}$, where $d$ is the period. The cavities are point defects formed by periodically removing air voids in a line with $D=2d$ (see Fig.~\ref{fig:schematic}~(c)). The complex frequency, $\tilde{\Omega}_0$, and the complex coupling parameter, $\tilde{\beta}_1$, of the structure are $(0.305-i7.71\times10^{-5})4\pi c/D$, and $9.87\times10^{-3}-i1.97\times10^{-5}$, respectively.\par
\begin{figure}
         
                \includegraphics[scale=0.7]{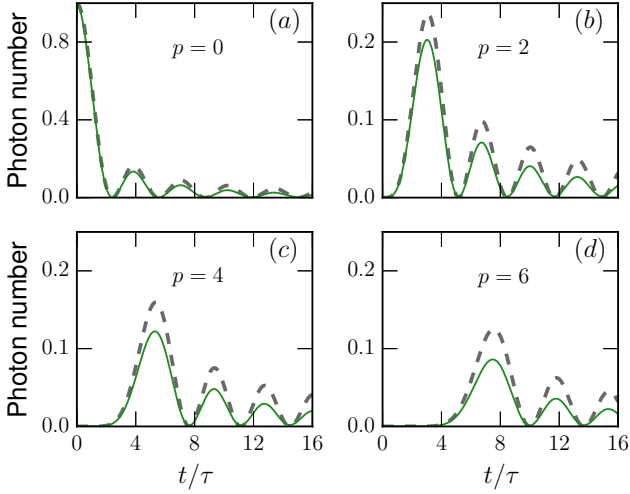}

                \caption{(Color online) Average photon number in the (a)~central, (b)~second, (c)~fourth, and (d)~sixth cavities as a function of time for the CROW structure. The dashed grey lines show the cases in which the effects of loss are ignored. Note the different scaling in (a).}
                \label{fig:Photon_number}
                
                \end{figure}

In Fig.~\ref{fig:Photon_number}, we plot the number of photons in the $p^{th}$ cavity for $p=0,2,4,6$ as a function of time for both lossy~(green) and lossless~(dashed grey) systems. As can be seen, due to the exponentially decaying term in Eq.~(\ref{eq:photon_number_CROW}), the number of photons in each cavity decreases as the system evolves in time. However, in addition to the cavity leakage into the environment, coupling between the cavities also affects the number of photons in each cavity. In the other words, even in the lossless system~(grey), we still see in Fig.~\ref{fig:Photon_number} that, due to the multiple photon hopping back and forth between the cavities, the maximum number of photons in the $p^{th}$ cavity gets smaller as $p$ gets larger. The propagation of light between the coupled cavities is evident in Figs.~\ref{fig:Photon_number}~(b-d). As can be seen, as the cavity index increases from $p=2$ to $p=6$, a longer time is needed for the photons to travel from the $c^{th}$ cavity to the $p^{th}$. If we consider a lossless system, then what one needs to calculate the time at which the photon number in each cavity reaches the maximum value is to find the first maximum of the Bessel function. From Ref.~\cite{Abramowitz:1974:HMF:1098650}, it can be shown that for the cavities far from the $c^{th}$ cavity~(large $p$), the first maximum occurs at the time
\begin{equation}
\frac{t_p}{\tau}\approx p+c_0 p^{1/3},
\end{equation} 
where $c_0\approx0.8$. Using Eqs.~(\ref{eq:dispersion_NNTB}) and (\ref{eq:photon_number_CROW}), the effective propagation velocity (defined as the distance to the cavity divided by the time required to reach the cavity) is then given approximately by
\begin{equation}
\label{eq:propagation_velocity}
v_p=\frac{pD}{t_p}\approx v_{max}(1-c_0 p^{-2/3}),
\end{equation}    
where $v_{max}=D/\tau$ is the maximum group velocity, which is about $0.04$ of the speed of light in vacuum for this CROW structure.  \par   
Summing the average number of photons in each cavity and applying Neumann's addition theorem~\cite{Olver:2010:NHM:1830479}, the total number of photons in the system at time $t$ is given by
\begin{equation}
N_{tot}=\sum_{p}{}\braket{a^\dagger _p (t) a_p(t)}=\braket{a^\dagger _c  a_{c}} e^{-2\gamma t} I_{0}\left( 2\Im \left(\tilde{\zeta}_1 t\right) \right),
\label{eq:total_photon_number_CROW}
\end{equation} 
where $I_0$ is the zeroth order modified Bessel function of the first kind. Ignoring the losses in the system, it can be shown that the total number of photons in the system is simply $\sinh[2](u)$ for all $t$, which is exactly the number of photons in the $c^{th}$ cavity at time $t=0$. Also, note that although $I_{0}\left(2\Im\left(\tilde{\zeta}_1 t\right)\right)$ monotonically increases with time, the total number of photons still decreases as the system evolves in time due to the dominant exponential factor, $e^{-2\gamma t}$, in Eq.~(\ref{eq:total_photon_number_CROW}). Using a series expansion of the Bessel function, it can be shown that for the time range considered in this paper, the total number of photons in the system is accurately given by
\begin{equation}
N_{tot}\approx \braket{a^\dagger _c  a_{c}} e^{-2\gamma t} \left(1+\left(\Im\left(\tilde{\zeta}_1 t\right)\right)^2\right),
\label{eq:N_correction}
\end{equation}
where the $\gamma$ is the decay constant associated with an individual cavity and the term involving $\Im\left(\tilde{\zeta}_1 t\right)$ is the correction term which arises due to the averaging of the losses over all the possible Bloch states. As mentioned earlier, different Bloch states experience different losses. Using Eq.~(\ref{eq:dispersion_NNTB}), it can be shown that the quality factor of the CROW mode at $k=0$ is $8.3$ times greater than mode with $k=\pi/D$. Thus we see that the decay in the total photon number is non-exponential.  
    
\begin{figure}
         
                \includegraphics[scale=0.7]{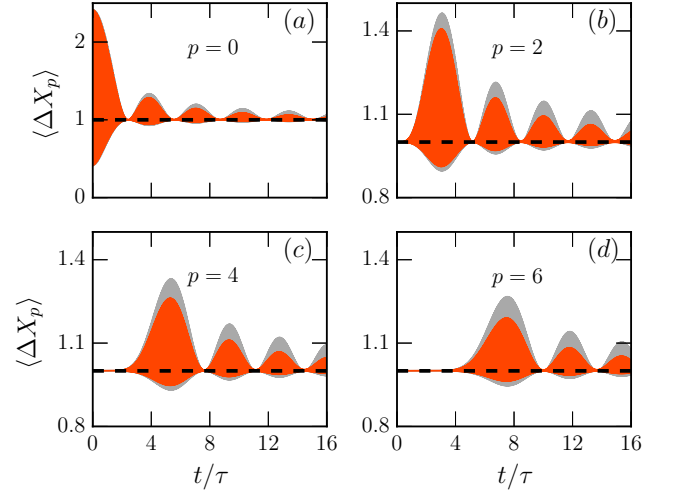}

                \caption{(Color online) Quadrature noise in $X$ in the (a)~central, (b)~second, (c)~fourth, and (d)~sixth cavities as a function of time for the CROW. The deviation from a lossless system, shown with light grey, is evident in each case. Note the different scaling in~(a) and note that there are fast oscillations that are not observable on this timescale.} 
                \label{fig:Delta_X_CROW}
                
                \end{figure}

Following a procedure similar to that used to arrive at Eqs.~(\ref{eq:photon_number_CROW}), one can derive the following expressions for the variances of the quadrature operators in the CROW structure:
\begin{equation}
\label{eq:DeltaX_CROW}
\begin{split}
\braket{(\Delta X_p)^2}=&1+2\braket{a^\dagger _c  a_{c}} e^{-2\gamma t} |J_{\delta p}(\tilde{\zeta}_1 t)|^2\\
&+\braket{a _c  a_{c}}e^{i\delta p \pi} J_{\delta p}^2(\tilde{\zeta}_1 t) e^{-2i\tilde{\Omega}_0 t}\\
&+\braket{a^\dagger _c  a^\dagger _{c}}e^{-i\delta p \pi} J_{\delta p}^2(\tilde{\zeta}^* _1 t) e^{2i\tilde{\Omega}^* _0 t}
\end{split}
\end{equation} 
and
\begin{equation}
\label{eq:DeltaY_CROW}
\begin{split}
\braket{(\Delta Y_p)^2}=&1+2\braket{a^\dagger _c  a_{c}} e^{-2\gamma t} |J_{\delta p}(\tilde{\zeta}_1 t)|^2\\
&-\braket{a _c  a_{c}}e^{i\delta p \pi} J_{\delta p}^2(\tilde{\zeta}_1 t) e^{-2i\tilde{\Omega}_0 t}\\
&-\braket{a^\dagger _c  a^\dagger _{c}}e^{-i\delta p \pi} J_{\delta p}^2(\tilde{\zeta}^* _1 t) e^{2i\tilde{\Omega}^* _0 t}.
\end{split}
\end{equation}
The time dependant quadrature noise in $X$ for lossy and lossless systems is shown in orange and grey, respectively, in Fig.~\ref{fig:Delta_X_CROW} for different cavities in the CROW structure. As can be seen in Fig.~\ref{fig:Delta_X_CROW}~(a), due to the nature of SVSs, the quadrature noise in $X$ is initially less than the classical limit (dashed line), as expected. The coupling between the cavities and the scattering loss in the system degrades the squeezing in the cavities as the system evolves in time. \par     
\begin{figure}
         
                \includegraphics[scale=0.44]{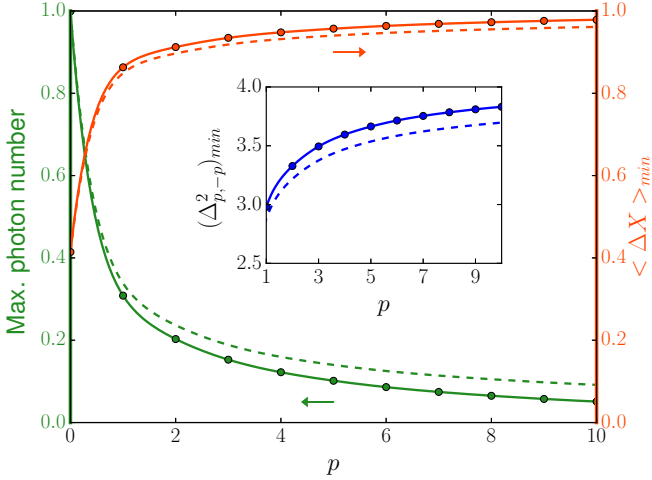}

                \caption{(Color online) Maximum number of photons (left axis) and minimum quadrature noise in $X$ (right axis) in the first eleven cavities of the CROW. The inset shows the minimum correlation variance between different symmetrically-displaced pairs of cavities. The dashed lines show the same quantities when the system is lossless.} 
                \label{fig:maximums}
                
                \end{figure}

The maximum number of photons and the maximum squeezing in each cavity is shown in Fig.~\ref{fig:maximums}. As expected, both the maximum number of photons and squeezing in the $X$ quadrature decrease as we move away from the central cavity. This is most evident when comparing the corresponding quantities for the tenth and the central cavities. The quadrature noise in $X$ and the maximum photon number in the tenth cavity are $2.4$ and $0.05$ times the corresponding values at the zeroth cavity, respectively. This figure also shows the effects of loss on the maximum number of photons and maximum squeezing by comparing the lossy~(solid line) and lossless~(dashed line) results. As can be seen, in the absence of loss, the maximum number of photons is higher, as expected, and the quadrature noise in $X$ is more squeezed, which is in agreement with our previous results on squeezed state generation in a single lossy cavity~\cite{Seifoory:17}. \par
We now study the inseparability criteria for the squeezed light inside the CROW structure. Using Eqs.~(\ref{eq:general_entanglement}), the time-dependent correlation variance can be written as
\begin{equation}
\begin{split}
\label{eq:Delta_nm}
\Delta_{p,p'}^2=&4+4\braket{a^\dagger _c  a_{c}}e^{-2\gamma t}\Big(|J_{\delta p}(\tilde{\zeta}_1 t)|^2+|J_{\delta p'}(\tilde{\zeta}_1 t)|^2\Big)\\
&-4\braket{a _c  a_{c}}e^{-2i\tilde{\Omega}_0 t}e^{i(\delta p+\delta p')\pi/2}J_{\delta p}(\tilde{\zeta}_1 t)J_{\delta p'}(\tilde{\zeta}_1 t)\\
&-4\braket{a^\dagger _c  a^\dagger _{c}}e^{2i\tilde{\Omega}^*_0 t}e^{-i(\delta p+\delta p')\pi/2}J_{\delta p}(\tilde{\zeta}^*_1 t)J_{\delta p'}(\tilde{\zeta}^*_1 t),
\end{split}
\end{equation}
where the sum of the last three terms needs to be negative for the inseparability criteria for CVs to be fulfilled.\par
\begin{figure}
         
                \includegraphics[scale=0.7]{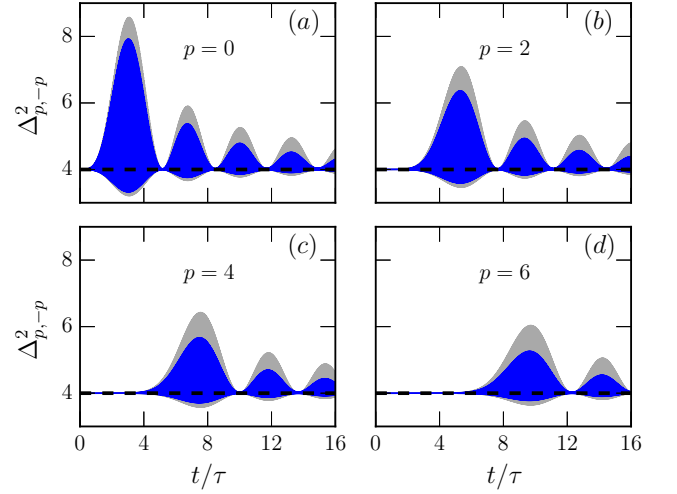}

                \caption{(Color online) Correlation variance between different pairs of cavities in the CROW as a function of time. The results for a lossless system are shown in light grey.} 
                \label{fig:different_n_m}
                
                \end{figure}

In Fig.~\ref{fig:different_n_m} we plot the time-dependant correlation variances for different sets of lossy and lossless cavities in blue and grey, respectively. The dashed lines show the inseparability criteria below which the light is considered to be entangled. Here we only focus on cases where the two cavities considered are located the same distance from the central cavity, as this will yield the maximum entanglement; however, using Eq.~(\ref{eq:Delta_nm}) one can explore the entanglement between any two cavities of the CROW. As can be seen, the maximum entanglement between each pair of cavities occurs when the peak in the photon number arrives at those cavities. Indeed, as time passes and the system evolves in time, the photons either scatter to the environment or move along the CROW, leading to a reduction in the number of photons in the considered cavity and consequently a decrease in the degree of entanglement. In order to compare the entanglement between different sets of cavities, in the inset to Fig.~\ref{fig:maximums} we plot the minimum correlation variance as a function of cavity index, $p$. The dashed lines show the results when the effects of loss are ignored. As expected, the difference between the lossy and the ideal systems is more evident as we get away from the central cavity. \par 

Finally, in order provide better insight into how the maximum number of photons, the minimum quadrature noise in $X$, and the minimum correlation variance varies with cavity number in a \textit{lossless} CROW system for the cavities far from the central cavity~(large $p$) when the $c=0$ cavity is initially in the SVS, we use asymptotic expansion expressions for the Bessel functions~\cite{Abramowitz:1974:HMF:1098650} in Eqs.~(\ref{eq:photon_number_CROW}), (\ref{eq:DeltaX_CROW}), and (\ref{eq:Delta_nm}) to obtain
\begin{equation}
\braket{a^\dagger _p a_p}_{max}\approx \frac{c_1^2 \sinh[2](u)}{p^{2/3}},
\end{equation} 
\begin{equation}
\braket{(\Delta X_p)^2}_{max}\approx 1-\frac{c_1^2(1-e^{-2u})}{p^{2/3}},
\end{equation}
and  
\begin{equation}
(\Delta_{p,-p}^2)_{max} \approx 4\left(1-\frac{c_1^2(1-e^{-2u})}{p^{2/3}}\right),
\end{equation}   
where $c_1=2^{1/3}Ai\left(-2^{1/3}c_0\right) \approx 0.67$, where $Ai$ is the Airy function. It can be seen that for all of these three quantities the dependance on $p$ is $p^{2/3}$. 
\section{Conclusion}
\label{sec:conclusion} 
In this work, we have examined the time evolution of squeezed states in coupled cavity systems. We have applied the tight-binding method to evaluate the fields and complex frequencies for the leaky modes of lossy coupled-cavity system.\par
We have presented the analytic time-dependant expressions for the photon number, quadrature noise, and correlation variance in the simple two coupled-cavity system and in a lossy CROW structure in terms of Bessel functions. \par
We have examined how the nonclassical properties of light in one cavity will be transferred to the other cavities in lossy coupled-cavity systems and have shown how loss affects properties such as photon number, quadrature squeezing, and entanglement. Moreover, we have studied the maximum values of these three quantities in both a lossy and a lossless system and have derived approximate analytic expression for the $p$-dependance of the maximum values of these three quantities in the absence of loss.\par 
We have found that for the CROW structure considered in this work, the effects of loss are significant and should not be neglected. These effects are most significant for cavities far from the excited cavity. The importance of loss will depend on the single-cavity loss, the group velocity and the loss dispersion of the particular CROW being studied.  Our analytic results allow for the investigation of these effects for any CROW that can be modelled using the nearest-neighbour tight-binding approximation.\par
Although we have focused on the squeezed states in this work, one can study the same quantities of the other states of light such as squeezed thermal states and coherent states by simply replacing the corresponding quantities in the expressions provided in this paper.\par
Finally, it should be mentioned that here, rather than employing a continuous-wave pump to generate nonclassical light in the central cavity, we have focused on the time evolution of initially generated squeezed state in the system. The full process of generation and evolution of squeezed state in a coupled-cavity system and evaluating the dynamics of the continuous variable entanglement in such a system under continuous-wave pumping will be explored in future work.  \par
\section{acknowledgement}
This work was supported by Queen's University and the Natural Sciences and Engineering Research Council of Canada~(NSERC). The authors would also like to gratefully thank John E. Sipe and Mohsen Kamandar Dezfouli for many fruitful discussions.        
\bibliography{mybib}
\end{document}